\documentstyle[12pt]{article}
\newcommand{\extraspace}{\addtolength{\abovedisplayskip}{2mm}
                        \addtolength{\belowdisplayskip}{2mm}
                        \addtolength{\abovedisplayshortskip}{2mm}
                        \addtolength{\belowdisplayshortskip}{2mm}}
\newcommand{\be}{\begin{equation}\extraspace}

\newcommand{\ee}{\end{equation}}

\newcommand{\bea}{\begin{eqnarray}\extraspace}
\newcommand{\beastar}{\begin{eqnarray*}\extraspace}
\newcommand{\eea}{\end{eqnarray}}
\newcommand{\eeastar}{\end{eqnarray*}}

\newcommand{\nonu}{\nonumber \\[2mm]}

\newcommand{\strutje}{\rule[-1.5mm]{0mm}{5mm}}
\newcommand{\str}{\rule[-2.5mm]{0mm}{7mm}}

\newcommand{\half}{\frac{1}{2}}
\newcommand{\quart}{\frac{1}{4}}

\newcommand{\si}{\sigma}
\newcommand{\la}{\lambda}

\newcommand{\cd}{c^{\dagger}}
\newcommand{\Sd}{S^{\dagger}}
\newcommand{\Sz}{S^z}
\newcommand{\etad}{\eta^{\dagger}}
\newcommand{\etaz}{\eta^z}
\newcommand{\Qd}{Q^{\dagger}}
\newcommand{\Qt}{\widetilde{Q}}
\newcommand{\Qtd}{\widetilde{Q}^{\dagger}}
\newcommand{\jp}{j^{\prime}}
\newcommand{\kp}{k^{\prime}}
\newcommand{\lp}{l^{\prime}}

\newcommand{\up}{\uparrow}
\newcommand{\down}{\downarrow}

\newcommand{\antivac}{\langle \, 0 \,|}
\newcommand{\vac}{| \, 0 \, \rangle}
\newcommand{\upstate}{|\!\! \up \, \rangle}
\newcommand{\downstate}{|\!\! \down \, \rangle}
\newcommand{\updownstate}{|\!\! \up\down \, \rangle}
\newcommand{\sumnn}{\sum_{\langle jk \rangle}}

\newcommand{\pr}{Phys.Rev.\ }
\newcommand{\prl}{Phys.Rev.Lett.\ }

\begin{document}
\baselineskip=15pt
\null\vskip -2cm
\hfill {ITP-SB-92-03, Feb.\ 1992}
\vskip 1.5cm
\begin{center}

{\Large New exactly solvable model of strongly correlated electrons}\\
\vspace{3mm}
{\Large motivated by high-$T_c$ superconductivity%
\footnote{Work supported in part by NSF grant PHY-9107261}}\\

\vskip 2.0cm

{\large Fabian H.L. E\char'31 ler%
\footnote{email: {\sc fabman@max.physics.sunysb.edu}} }\\
\vspace{2mm}
{\large Vladimir E. Korepin%
\footnote{email: {\sc korepin@dirac.physics.sunysb.edu}} }\\
\vspace{2mm}
{\large  and}\\
\vspace{2mm}
{\large Kareljan Schoutens%
\footnote{email: {\sc schoutens@sunysbnp.bitnet}} }

\vskip .5cm

{\sl Institute for Theoretical Physics\\
     State University of New York at Stony Brook\\
     Stony Brook, NY 11794-3840, U.S.A.}

\vskip 1.5cm

{\bf Abstract}

\end{center}

\vspace{.3cm}

\baselineskip=18pt

\noindent

We present a new model describing strongly correlated electrons
on a general $d$-dimensional lattice. It differs from the Hubbard model
by interactions of nearest neighbours, and it contains
the $t$-$J$ model as a special case.

The model naturally describes local electron pairs, which can move
coherently at arbitrary momentum. By using an $\eta$-pairing mechanism
we can construct eigenstates of the
hamiltonian with off-diagonal-long-range-order (ODLRO). These might help
to relate the model to high-$T_c$ superconductivity.

On a one-dimensional lattice, the model is exactly solvable by Bethe Ansatz.

\vfill

\noindent PACS \ \
71.20.Ad\ \  % Electron states: developments in mathematical
             % and computational techniques
75.10.Jm\ \  % Magnetic properties and materials: Quantised spin models

\newpage

\baselineskip=18pt

The study of strongly correlated elctrons on a lattice is
an important tool in theoretical condensed matter physics
in general, and in the study of high-$T_c$ superconductivity
in particular. Two well-studied models are the Hubbard model
and the $t$-$J$ model. On a one-dimensional lattice
these models are both exactly solvable by Bethe Ansatz. In this
letter we propose a new model, which is again solvable in one
dimension, and which combines and extends some of the interesting
features of the Hubbard model and the $t$-$J$ model.

\vskip 6mm

Electrons on a lattice are described by operators $c_{j,\sigma}\ $,
$j=1,\ldots,L$, $\sigma=\up,\down$, where $L$ is the total number
of lattice sites. These are canonical Fermi operators with anti-commutation
relations given by $\{ c^\dagger_{i,\sigma} , c_{j,\tau} \}
= \delta_{i,j} \delta_{\sigma,\tau}$. The state
$\vac $ (the Fock vacuum) satisfies $c_{i,\sigma} \vac = 0$.
At a given lattice site $i$ there are four possible electronic states
\be
\vac \ , \qquad
\upstate_i = \cd_{i,\up} \vac \ , \qquad
\downstate_i = \cd_{i,\down} \vac \ , \qquad
\updownstate_i = \cd_{i,\down} \cd_{i,\up} \vac \ .
\label{states}
\ee
By $n_{i,\si}= c^\dagger_{i,\si} c_{i,\si}$ we denote the number
operator for electrons with spin $\si$ on site $i$ and we write
$n_i=n_{i,\up} + n_{i,\down}$. The spin-operators $S=\sum_{j=1}^L
S_j\, , \quad$, $\Sd\, , \quad$ $S^z$,
\be
S_j = \cd_{j,\up} c_{j,\down}\ , \qquad
\Sd_j = \cd_{j,\down} c_{j,\up}\ ,\qquad
\Sz_j = \half (n_{j,\up}-n_{j,\down}) \ ,
\label{su2spin}
\ee
form an $SU(2)$ algebra and they commute with the hamiltonians
that we consider below. [We shall always give local expressions
${\cal O}_j$ for symmetry generators, implying that the global ones
are obtained as ${\cal O} = \sum_{j=1}^L {\cal O}_j$.]

The Hubbard model hamiltonian can be written as
\be
H^{\rm Hubbard} = - \sumnn \sum_{\sigma=\up,\down}
    (\cd_{j,\si} c_{k,\si} + \cd_{k,\si} c_{j,\si}) +
    U \sum_{j=1}^L (n_{j,\up}-\half)(n_{j,\down}-\half) \,,
\label{huham}
\ee
where the first summation runs over all nearest neighbour pairs
$\langle jk \rangle$.
It contains kinetic (hopping) terms for the electrons and an
on-site interaction term for electron pairs. An interesting feature
(on a bipartite periodic lattice) is the so-called $\eta$-pairing symmetry
\cite{eta,pairing}, which involves operators
$\eta_H$, $\etad_H$ and $\etaz_H$ which form another $SU(2)$ algebra,
and which commute with the hamiltonian (\ref{huham}). Using this symmetry
one can, starting from an eigenstate $|\psi\rangle$ of the hamiltonian,
create a new eigenstate $\etad_H |\psi\rangle$, which contains an additional
local electron pair of momentum $\pi$.
The spin $SU(2)$ algebra (\ref{su2spin}) and the $\eta$-pairing $SU(2)$
algebras together form an $SO(4)$ symmetry algebra.
In one dimension, the Hubbard model is solvable by Bethe Ansatz \cite{liebwu}.

In the $t$-$J$ model, there is a kinematical constraint which forbids
the occurence of two electrons on the same lattice site. On this restricted
Hilbert space the $t$-$J$ hamiltonian (with $t=1$, $J=2$) acts as
$H^{t-J} = - \sumnn H^{t-J}_{j,k}$, with
\bea
H^{t-J}_{j,k} &=& \sum_{\sigma=\up,\down}
    (\Qd_{j,\si} Q_{k,\si} + \Qd_{k,\si} Q_{j,\si})
\nonu
&&  - 2 \left( S^z_j S^z_{k} + \half (\Sd_j S_k + S_j \Sd_k)
    - \half (1-n_j-n_k) - \quart n_j n_k \right) \ ,
\label{tJham}
\eea
where we defined
\be
Q_{j,\up} = (1-n_{j,\down})\, c_{j,\up}\ , \qquad
Q_{j,\down} = (1-n_{j,\up})\, c_{j,\down}
\label{susy}
\ee
and the operators $S^z_j$, $S_j$ and $\Sd_j$ are as in (\ref{su2spin}).
The $t$-$J$ model (\ref{tJham}) is supersymmetric and the spin $SU(2)$
symmetry algebra gets enlarged to the superalgebra $SU(1|2)$
\cite{BBO,sarkar} (see \cite{corn} for the description and classification
of the classical Lie superalgebras). The generators of this symmetry algebra
are $S$, $\Sd$, $S^z$, $Q_{\up}$, $\Qd_{\up}$, $Q_{\down}$, $\Qd_{\down}$
and $T= 2 L - \sum_{j=1}^L n_j$.
In one dimension the supersymmetric $t$-$J$ model (\ref{tJham})
is exactly solvable by Bethe Ansatz \cite{lai,suth,schlott}.

\vskip 6mm

Before we present the hamiltonian of the new model, we give some
motivation, which is based on what we know about the materials that
exhibit high-$T_c$ superconductivity. It has been found that the
electrons in these materials form `Cooper pairs', which are
spin-singlets, and that these pairs are much smaller than in the
traditional superconductors. As a limiting case one can consider
models which have electron pairs of size zero, {\it i.e.,} pairs
that are localised on single lattice sites. We will call such
localised electron pairs {\it localons}.

In the $t$-$J$ model localons are ruled out by the kinematical
constraint on the space of states, and in the Hubbard model only
local pairs of momentum $\pi$ exist. Below we shall see that
in our new model localons can move coherently with arbitrary momentum.
Apart from these local pairs, the new model may also have
bound states that are finite-size electron pairs.

\vskip 6mm

Let us now present the hamiltonian of the new model on a general
$d$-dimensional lattice. We write it as
\be
H = H^0
      + U \, \sum_{j=1}^L (n_{j,\up}-\half)(n_{j,\down}-\half)
      + \mu \, \sum_{j=1}^L n_j
      + h \, \sum_{j=1}^L (n_{j,\up}-n_{j,\down}) \ ,
\label{hamil}
\ee
where $H^0$ is given by
\be
H^0 = - \sumnn \, H^0_{j,k} \ ,
\qquad \langle j,k \rangle\ {\rm are}\ {\rm n.n.}
\label{hamil0}
\ee
with
\bea
H^0_{j,k} &=&
   \cd_{k,\up} c_{j,\up}(1-n_{j,\down}-n_{k,\down})
   + \cd_{j,\up} c_{k,\up}(1-n_{j,\down}-n_{k,\down})
\nonu
&& + \cd_{k,\down} c_{j,\down}(1-n_{j,\up}-n_{k,\up})
   + \cd_{j,\down} c_{k,\down}(1-n_{j,\up}-n_{k,\up})
\nonu
&& + \half (n_j - 1)(n_k - 1)
   + \cd_{j,\up} \cd_{j,\down} c_{k,\down} c_{k,\up}
   + c_{j,\down} c_{j,\up} \cd_{k,\up} \cd_{k,\down}
\nonu
&& - \half (n_{j,\up}-n_{j,\down})(n_{k,\up}-n_{k,\down})
   - \cd_{j,\down} c_{j,\up} \cd_{k,\up} c_{k,\down}
   - \cd_{j,\up} c_{j,\down} \cd_{k,\down} c_{k,\up}
\nonu
&& + (n_{j,\up}-\half)(n_{j,\down}-\half)
              + (n_{k,\up}-\half)(n_{k,\down}-\half) \ .
\label{hamil0jk}
\eea
This hamiltonian contains kinetic terms and interaction terms that
combine those of the Hubbard and of the $t$-$J$ model.
The second term in (\ref{hamil})
is the on-site Hubbard interaction term (notice that it also gets a
contribution from $H^0$). The third and the fourth terms in
(\ref{hamil}) introduce a chemical potential $\mu$ and a magnetic field $h$.
Roughly speaking, the new model can be viewed as a modified Hubbard model
with additional nearest neighbour interactions similar to those in the
$t$-$J$ model.

The hamiltonian $H^0$ is
invariant under spin reflection $c_{j,\up} \leftrightarrow
c_{j,\down}$ and under particle-hole replacement
$\cd_{j,\si} \leftrightarrow c_{j,\si}$. In addition to the
spin $SU(2)$ generators (\ref{su2spin}), the following operators
commute with $H^0$
\begin{itemize}
\item
$\eta$-pairing $SU(2)$, with generators $\eta$, $\etad$, $\etaz$,
\be
\eta_j = c_{j,\up} c_{j,\down}\ , \qquad
\etad_j = \cd_{j,\down} \cd_{j,\up}\ , \qquad
\etaz_j = - \half n_j + \frac{1}{2}\ .
\label{su2pair}
\ee
Together with the spin $SU(2)$ algebra (\ref{su2spin}), this gives
an $SO(4)$ algebra which is similar to the one for the Hubbard model.
This symmetry makes it possible to generalise the $\eta$-pairing
mechanism, which was developed for the Hubbard model in \cite{pairing},
to the new model.
\item
Supersymmetries. There are eight supersymmetries in total:
$Q_{\up}$, $\Qd_{\up}$, $Q_{\down}$ and $\Qd_{\down}\ $ given in
(\ref{susy}) and the operators $\Qt_{\si}$, $\Qtd_{\si}$,
\be
\Qt_{j,\up} =  n_{j,\down} c_{j,\up}\ , \quad
\Qt_{j,\down} = n_{j,\up} c_{j,\down}\ .
\label{susytilde}
\ee
\end{itemize}
These generators, together with the operator $\sum_{j=1}^L 1$
(which is constant and equal to $L$), form the superalgebra $SU(2|2)$.
[Like $SU(4)$, this algebra has 15 generators, eight of which are
fermionic. In the fundamental representation, the generators can be
represented as $4\times 4$ supermatrices with vanishing supertrace
\cite{corn}.]

The symmetries of the hamiltonian $H^0$ can be made manifest as
follows. We first add one more generator to the symmetry algebra,
which is
\be
X=\sum_{j=1}^L X_j\ , \qquad
X_j = (n_{j,\up}-\half)(n_{j,\down}-\half)\ , \qquad
[H^0,X] = 0\ .
\label{csaX}
\ee
This extends the superalgebra $SU(2|2)$ to $U(2|2)$. We denote the
generators of this algebra by $J_\alpha$, where $\alpha=1,2,\ldots,16$.
We now introduce an invariant, non-degenerate 2-index tensor,
denoted by $K^{\alpha\beta}$, which is the inverse of
$K_{\alpha\beta} = str(J_{\alpha} J_{\beta})$, where the
$J_{\alpha}$ are $4\times 4$ supermatrices in the fundamental
representation. Using this, we can cast $H^0_{j,k}$ in a group-theoretical
form as follows
\bea
H^0_{j,k} &=&
   \sum_{\alpha,\beta=1}^{16} K^{\alpha\beta} J_{j,\alpha} J_{k,\beta}
\label{group}
\\
&=&
\sum_{\sigma=\up,\down}
   \left( \Qd_{j,\si} Q_{k,\si} + \Qd_{k,\si} Q_{j,\si}
          - \Qtd_{j,\si} \Qt_{k,\si} - \Qtd_{k,\si} \Qt_{j,\si} \right)
\nonu
&& + (2 \, \etaz_j \etaz_k
   + \etad_j \eta_k + \eta_j \etad_k)
   - (2 \, \Sz_j \Sz_k
   + \Sd_j S_k + S_j \Sd_k) + X_j + X_k \ .
\eea
It is easily checked that this expression agrees with
the formula for $H^0_{j,k}$ in (\ref{hamil0jk}). The expression
(\ref{group}) immediately makes it clear that $H^0$ commutes with all
16 generators of $U(2|2)$.

We would like to stress that the appearance of the algebra $U(2|2)$
in the model is not too surprising: on each lattice site there are
four electronic states (\ref{states}), two of which are fermionic.
The supergroup $U(2|2)$ is the group of all unitary rotations of
these four states into one another. Our hamiltonian $H^0$
has been chosen such that it commutes
with the entire algebra $U(2|2)$ and is therefore very natural. The
analogous construction for $U(1|2)$ leads to the
$t$-$J$ hamiltonian (\ref{tJham}), and for $U(2)$ it leads to the
spin-$\half$ $XXX$ Heisenberg model.

The spectrum of the hamiltonian $H^0$ is symmetric around zero.
This follows from the discrete symmetry
$\cd_{j,\down} \leftrightarrow c_{j,\down}$, for which
$H^0 \leftrightarrow -H^0 $.

There is a further aspect of $H^0$ that deserves to be
mentioned: the terms $H^0_{j,k}$ act as a graded permutations
of the electron states (\ref{states}) at sites $j$ and $k$.
By `graded' we mean that there is an extra minus sign if the
two states that are permuted are both (fermionic) single electron
states. For example,
\be
H^0_{j,k} \, \cd_{j,\up} \vac
    = \cd_{k,\up} \vac \, , \quad
H^0_{j,k} \, \cd_{j,\up} \cd_{k,\down} \vac
    = - \cd_{j,\down} \cd_{k,\up} \vac \, , \quad
{\rm etc.}
\label{perm}
\ee
In this respect, the new model generalises the spin-$\half$ $XXX$
model and the $t$-$J$ model (\ref{tJham}).
The nearest neighbour hamiltonians of these models have a similar
interpretation as graded permutations of the basic states, which are
$\{ \upstate,\downstate \}$ for the spin-$\half$ $XXX$ model and
$\{ \vac,\upstate,\downstate \}$ for the $t$-$J$ model.
Lattice hamiltonians that act like (graded) permutations were first
considered by Sutherland in \cite{suth}.

We define the number operators $N_\up$, $N_\down$ (the number of
single electrons with given spin) and $N_l$ (the number of localons)
by
\be
N_\up +N_l = \sum_{j=1}^L n_{j,\up}\,, \qquad
N_\down + N_l = \sum_{j=1}^L n_{j,\down}\,, \qquad
N_l = \sum_{j=1}^L n_{j,\up}\ n_{j,\down}
\label{nums}
\ee
and we write $N_e=N_\up+N_\down$ for the total number of single
electrons. The fact that $H^0$ is a permutation
makes it clear that these number operators commute with $H^0$, so that
$H^0$ can be diagonalised within a sector with given numbers $N_{\up}$,
$N_{\down}$ and $N_l$. This implies that the terms proportional to $U$,
$\mu$ and $h$ in (\ref{hamil}), which break the symmetry $U(2|2)$,
will not affect the solvability of the model in one dimension.

In the sectors without localons $H^0$ reduces to the $t$-$J$
hamiltonian (\ref{tJham}). (This is clear from the fact that they both
act as permutations.) The new model reduces to the spin-$\half$
$XXX$ model in the sector with only vacancies and localons, and similarly
in the (half-filled) sector with one single electron on each site.

\vskip 6mm

Let us now briefly discuss some physical aspects of the new model.
We first remark that we can always (for general lattices in an
arbitrary number of dimensions) construct a number of exact
eigenstates of the hamiltonian which show
off-diagonal-long-range-order (ODLRO), which is characteristic for
superconductivity \cite{odlro}. For this
we follow the construction which was developed for the Hubbard model
by C.N. Yang in \cite{pairing}. The state $ \Psi_N = (\etad)^N \vac\ $
is an eigenstate of the hamiltonian with energy
$ E= 2 \mu N + U \frac{L}{4} - M $,
where $M$ is the total number of nearest neigbour links
$\langle jk \rangle$ in the lattice. Following \cite{pairing},
we compute the following off-diagonal matrix element ($k\neq l$)
of the reduced density matrix $\rho_2$
\be
\langle (k,\down) (k,\up) | \, \rho_2 \, | (l,\up) (l,\down) \rangle =
\frac{ \strutje \antivac \eta^N \cd_{k,\down} \cd_{k,\up}
   c_{l,\up} c_{l,\down} (\etad)^N \vac }{ \str
   \antivac \eta^N (\etad)^N \vac}
       = \frac{N(L-N)}{L(L-1)} \ .
\ee
The fact that this off-diagonal matrix element is constant for large
distances $|j-k|$ establishes the property of ODLRO for the states $\Psi_N$.
In the thermodynamic limit we have ODLRO as soon as the occupation number
$N$ of the zero-momentum localon state becomes macroscopic, {\it i.e.}, when
$ N/L \rightarrow D $ becomes finite.

The local electron pairs that participate in the $\eta$-pairing
have momentum zero. However, the model also admits localons that move
with arbitrary momentum. This follows from the fact that $H^0_{j,k}$
acts as a permutation of the electronic states (\ref{states}) on
neighbouring sites: due to this
localons cannot decay and move coherently. On a $d$-dimensional square
lattice (with lattice spacing $a$) the wavefunction
$ \sum_{\vec{x}} \, \exp (i \vec{x} \vec{k}) \,\,
  \cd_{\vec{x},\up} \cd_{\vec{x},\down} \vac $,
which describes a single localon of momentum $\vec{k}$ over the
bare vacuum, is an exact eigenfunction of the hamiltonian
(\ref{hamil}) of energy
\be
  E= 2d - 2 \sum_{m=1}^d \cos(k_m a)
  + U \frac{L}{4} + 2\mu -M \ .
\ee

Multi-localon wave-functions, as well as wavefunctions with
single electrons, exist but cannot easily be written down for
higher-dimensional lattices.
However, in one dimension the model is exactly solvable by
Bethe Ansatz (BA), and we can obtain explicit expressions for general
eigenstates of the hamiltonian. We think that it is worthwhile to
study this exact solution, and that this will lead to a better
understanding of the higher-dimensional model as well.

We will here briefly summarise the results of the exact solution
in one dimension; the details are deferred to a separate publication
\cite{tbp}. The exact solution starts from the
observation that the hamiltonian is a graded permutation (\ref{perm})
of four states, of which two are fermionic and two are bosonic.
The BA analysis for hamiltonians which are graded permutations was
first considered by Sutherland in \cite{suth}, see also \cite{kulish}.
The method of solution is the algebraic version of the `nested Bethe
Ansatz' \cite{nestedBA} (for an introduction to the algebraic BA,
see \cite{book}). Each step of the nesting involves the
introduction of a set of spectral parameters, which are in our case
$\la_j$, $\la^{(1)}_k$ and $\la^{(2)}_l$, where
$j=1,\ldots,(N_e+N_l)$, $k=1,\ldots, N_e$ and $l=1,\ldots, N_{\down}$.

For each choice of a set of rapidities we can construct
an eigenstate of the hamiltonian $H^{0}$ in the sector specified
by $N_l$, $N_e$ and $N_{\down}$, with energy $E^0$ given by
\be
E^0 =  \sum_{j=1}^{N_e+N_l} \frac{1}{\la_j^2+\frac{1}{4}} - L \ .
\label{energy}
\ee

The boundary conditions for these eigenstates lead to the following
set of Bethe equations for the rapidities $\la_j$, $\la^{(1)}_k$ and
$\la^{(2)}_l$
\bea
\left(\la_j-{i\over 2}\over \la_j+{i\over 2}\right)^L &=&
\prod_{\scriptstyle \jp =1\atop \strut\scriptstyle \jp \neq j}^{N_e+N_l}
{\la_j-\la_{\jp} -i\over\la_j - \la_{\jp} +i}\
\prod_{\kp =1}^{N_e}{\la_{\kp}^{(1)}-\la_j -{i\over 2} \over
\la_{\kp}^{(1)}-\la_j+{i\over 2}}
% \, , \quad j=1\ldots N_e+N_l
\nonu
\prod_{\jp=1}^{N_e+N_l}{\la_k^{(1)}-\la_{\jp} +{i\over 2} \over
\la_k^{(1)}-\la_{\jp}-{i\over 2}} &=&
\prod_{\lp=1}^{N_{\down}}
{\la_k^{(1)}-\la_{\lp}^{(2)} +{i\over 2} \over
\la_k^{(1)}-\la_{\lp}^{(2)} -{i\over 2}}
% \, , \quad k=1\ldots N_e
\nonu
\prod_{\scriptstyle \lp =1\atop \strut\scriptstyle \lp\neq l}^{N_{\down}}
{\la_{\lp}^{(2)}-\la_l^{(2)} +i \over {\la_{\lp}^{(2)}-\la_l^{(2)} -i}} &=&
\prod_{\kp=1}^{N_e}
{\la_{\kp}^{(1)}-\la_l^{(2)} +{i\over 2} \over
\la_{\kp}^{(1)}-\la_l^{(2)}-{i\over 2}}\, .
% \quad l=1\ldots N_{\down}\ .
\eea
These equations, together with the expression (\ref{energy})
for the energy, guarantee that we shall be able to describe
explicitly the ground state of our model for arbitrary density
of electons, coupling constant $U$ and magnetic field $h$
\cite{tbp}.


\frenchspacing
\begin{thebibliography}{11}

\bibitem{eta}
  O.J. Heilmann and E.H. Lieb, Annals of the New York Acad. of Sci.
  {\bf 172} (1971) 583; E.H. Lieb, Phys.Rev.Lett. {\bf 62} (1989)
  1201;
\bibitem{pairing}
  C.N. Yang, \prl {\bf 63} (1989) 2144; \hfill\break
  C.N. Yang and S. Zhang, Mod.Phys.Lett. {\bf B4} (1990) 759
\bibitem{liebwu}
  E. Lieb and F.Y. Wu, Phys.Rev.Lett. {\bf 20} (1968) 1445
\bibitem{BBO}
  P.-A. Bares, G. Blatter and M. Ogata, \pr {\bf B44} (1991) 130
\bibitem{sarkar}
  S. Sarkar, J.Phys. {\bf A23} (1990) L409; J.Phys. {\bf A24} (1991)
  1137; J.Phys. {\bf A24} (1991) 5775
\bibitem{corn}
  J.F. Cornwell, {\it Group Theory in Physics, Vol III: Supersymmetries
  and Infinite-Dimensional Algebras}, Academic Press (1989). The
  algebras $SU(n|m)$ are discussed on page 270.
\bibitem{lai}
  C.K. Lai, J.Math.Phys. {\bf 15} (1974) 167
\bibitem{suth}
  B. Sutherland, \pr {\bf B12} (1975) 3795
\bibitem{schlott}
  P. Schlottmann, \pr {\bf B36} (1987) 5177
\bibitem{odlro}
  C.N. Yang, Rev.Mod.Phys. {\bf 34} (1962) 694
\bibitem{tbp}
  F.H.L. E\char'31 ler, V.E. Korepin and K. Schoutens,
  in preparation
\bibitem{kulish}
  P.P. Kulish, Jour.Sov.Math. {\bf 35} (1986) 2644;\hfill\break
  C.L. Schultz, Physica {\bf 122A} (1983) 71
\bibitem{nestedBA}
  C.N. Yang, Phys.Rev.Lett. {\bf 19} (1967) 1315
\bibitem{book}
  V.E. Korepin, G. Izergin and N.M. Bogoliubov,
  {\it Quantum Inverse Scattering Method, Correlation Functions
  and Algebraic Bethe Ansatz}, Cambridge University Press, 1992
\end{thebibliography}
\end{document}